\documentclass{emulateapj}

\shorttitle{SDSS $\times$ FIRST Quasar Luminosity Evolution and Radio Loudness}
\shortauthors{Singal et al.}

\slugcomment{accepted to ApJ}

\begin{document}
\title{THE RADIO AND OPTICAL LUMINOSITY EVOLUTION OF QUASARS II - THE SDSS SAMPLE}

\author{J. Singal\altaffilmark{1}, V. Petrosian\altaffilmark{1}$^,$\altaffilmark{2}, {\L}. Stawarz\altaffilmark{3}$^,$\altaffilmark{4}, A. Lawrence\altaffilmark{5}}

\altaffiltext{1}{Kavli Institute for Particle Astrophysics and Cosmology\\SLAC National Accelerator Laboratory and Stanford University\\382 Via Pueblo Mall, Stanford, CA 94305-4060}
\altaffiltext{2}{Also Departments of Physics and Applied Physics}
\altaffiltext{3}{Institute of Space and Astronautical Science (ISAS)\\ Japan Aerospace Exploration Agency (JAXA),\\ 
3-1-1 Yoshinodai, Chuo-ku, Sagamihara, Kanagawa 252-5510 Japan}
\altaffiltext{4}{Astronomical observatory of the Jagiellonian University\\ ul. Orla 171, 30-244 Krak\'ow, Poland}
\altaffiltext{5}{University of Edinburgh Institute for Astronomy\\Scottish Universities Physics Alliance (SUPA)\\Royal Observatory, Blackford Hill, Edinburgh UK}

\email{jsingal@stanford.edu}

\begin{abstract}
We determine the radio and optical luminosity evolutions and the true distribution of the radio loudness parameter $R$, defined as the ratio of the radio to optical luminosity, for a set of more than 5000 quasars combining SDSS optical and FIRST radio data.  We apply the method of Efron and Petrosian to access the intrinsic distribution parameters, taking into account the truncations and correlations inherent in the data.  We find that the population exhibits strong positive evolution with redshift in both wavebands, with somewhat greater radio evolution than optical.  With the luminosity evolutions accounted for, we determine the density evolutions and local radio and optical luminosity functions.  The intrinsic distribution of the radio loudness parameter $R$ is found to be quite different than the observed one, and is smooth with no evidence of a bi-modality in radio loudness for $\log R \geq -1$.  The results we find are in general agreement with the previous analysis of Singal et al. 2011 which used POSS-I optical and FIRST radio data.
\end{abstract}

\section{Introduction} \label{intro}

Quasars are distant active galactic nuclei (AGN) with emission seen across the electromagnetic spectrum, from radio to X-rays.  The optical emission from quasars is thought to be dominated by the radiation from the accretion disk around supermassive black holes, while the radio emission is dominated by the plasma outflowing from the black hole/accretion disk systems.  Because of this, different but complementary information can be gathered from both photon energy ranges regarding the cosmological evolution of AGN.  It is therefore important to determine in detail the redshift evolutions of quasars in both radio and optical  regimes.

In a previous paper \citep[hereafter QP1]{QP1} we explored the luminosity evolutions and radio loudness distribution of quasars with a dataset consisting of the overlap of the FIRST (Faint Images of the Radio Sky at Twenty Centimeters) radio survey with the Automatic Plate Measuring Facility catalog of the Palomar Observatory Sky Survey (POSS-I), as presented by \citet{White00}.  In this paper we present the results from a much larger dataset consisting of the overlap of FIRST with the Sloan Digital Sky Survey (SDSS) Data Release 7 \citep{SDSSDR7} quasar catalog, which contains a factor of $\sim 10$ more objects and spans redshifts from 0.065 to 5.46.  

In the literature, the evolution of the quasar luminosity function (LF) has been described not only for optical and radio luminosities but also for X-ray, infrared, and bolometric luminosities \citep[e.g.][]{Ueda03,R06,Matute06,Hopkins07,Croom09}.  The shape of the LF and its evolution are usually obtained from a flux limited sample in the $a$ waveband $f_a > f_{m,a}$ with $f_{m,a}$ denoting the flux limit and $L_a = 4 \, \pi \, d_L^2 f_a / K_a$, where $d_L(z)$ is the luminosity distance and $K_a(z)$ stands for the K-correction.  For a pure power law emission spectrum of index $\varepsilon_a$ defined as $f_a \propto \nu^{-\varepsilon_a}$, one has $K_a(z) = (1+z)^{1-\varepsilon_a}$.  This simple form may be modified for optical data by the presence of emission lines, as in this work.  

In general, the determination of the LF and its evolution requires analysis of the bi-variate distribution $\Psi_a(L_a,z)$.  The first step of the process should be the determination of whether the variables of the distributions are correlated or are statistically independent.  A correlation between $L_a$ and $z$ is a consequence of luminosity evolution.  In the case of quasars with the optical and some other band luminosity, we have at least a tri-variate function. One must determine not only the correlations between the redshift and individual luminosities (i.e. the two luminosity evolutions) but also the possible intrinsic correlation between the two luminosities, before individual distributions can be determined.  Knowledge of these correlations and distributions are essential for not only constraining robustly the cosmological evolution of active galaxies, but also for interpretation of related observations, such as the extragalactic background radiation \citep[e.g.][]{Singal10,Hopkins10}. 

A related question is the distribution of the `radio-loudness parameter', $R=L_{\rm rad}/L_{\rm opt}$, for the quasar population, defined as the ratio of the 5 GHz radio to 2500 \AA \, optical luminosity spectral densities, and the distinction between so-called `radio loud' ($R>10$; RL for short) and `radio quiet' ($R<10$; RQ for short) quasars.  Weak hints of a bi-modality in the distribution of the radio loundess parameter described by \citet{Kellerman89} suggested that $\log R = 1$ could be chosen as the radio loud/quiet demarcation value.  Using this value for the division between RL and RQ quasars, the differences between the two classes have been investigated, including the possibility of distinct cosmological evolution of the RL and RL populations \citep[e.g.][]{Miller90,Goldschmidt99,Jiang07}.   Still, the more recent analyses of different samples of objects reported in the literature so far gave rather inconclusive results on whether any bi-modality in the distribution of the radio loudness parameter for quasars is inherent in the population \citep[see][]{Ivezic02,Cira03,Ivezic04,Zamfir08,Kimball11,Mahony12}.  In QP1 we found no evidence for a bi-modality in $R$ in the range $-1<\log R<4$.

In addition to the above cited works, there have been many papers dealing with this ratio and RL vs RQ issue, as well as luminosity ratios at other wavelengths, e.g. IR/radio, Optical/X-ray etc.  However, to the best of our knowledge, except for QP1, none of these works have  address the correlations between the radio and optical luminosities, which is necessary for such and analysis.   Additionally, in general they have not concentrated on obtaining the {\it intrinsic} --- as opposed to the raw observed --- distribution (and/or evolution) of the ratio, which is related to the tri-variate LF $\Psi (L_{\rm opt},  L_{\rm rad}, z)$ by\footnote{Equation \ref{Gtrue} arises because by definition $\int {G_R(R,z) \, dR} = \int \int {\Psi(L_{\rm opt}, \, L_{\rm rad} \, z) \, dL_{\rm opt} \, dL_{\rm rad}}$, and, following from the definition of $R$, $dL_{\rm rad} = L_{\rm opt} \, dR$ and $dL_{\rm opt} = -(L_{\rm rad} / R^2) \, dR$.}
\begin{eqnarray}
G_R(R,z) = \int_0^{\infty} { \Psi (L_{\rm opt},  R \,\, L_{\rm opt}, z) \, L_{\rm opt} \, dL_{\rm opt} } \nonumber \\ = \int_0^{\infty} { \Psi \left( {L_{\rm rad} \over R},  L_{\rm rad}, z \right) \, L_{\rm rad} \, {{ dL_{\rm rad} } \over {R^2}} } .
\label{Gtrue}
\end{eqnarray}

In Appendix A of QP1 we showed how, even in the simplest cases, the observed distributions can be very different from the intrinsic ones.  Thus, for determination of the true distributions the data truncations must be determined and the correlations between all variables must be properly evaluated. 

\citet[EP for short]{EP92,EP99} developed  new methods for determination of the existence of correlation or independence of  variables from a flux limited and more generally truncated dataset, which were further expanded in QP1.  Our aim in this paper is to take all the selection and correlation effects into account in determination of the true evolution of optical and radio luminosities and their ratio and to find their distributions, using the larger SDSS DR7 QSO $\times$ FIRST dataset.   

In \S \ref{datasec} we describe the radio and optical data used.  \S \ref{simlumf} contains a general discussion of luminosity evolution and the sequence of the analysis.  In \S \ref{evsec} we apply the EP method to achieve the luminosity-redshift evolutions and the correlation between the luminosities.  We determine the density evolution in \S \ref{dev}, and the local luminosity functions in \S \ref{local}.  A discussion of the radio loudness distribution is presented in \S \ref{Rdist}.  \S \ref{testass} we investigate some of the assumptions used, and \S \ref{disc} contains a discussion of the results.  This work assumes the standard $\Lambda$CDM cosmology throughout, with $H_0=71$\,km\,s$^{-1}$\,Mpc$^{-1}$, $\Omega_{\Lambda}=0.7$ and $\Omega_{m}=0.3$.

\section{Data} \label{datasec}

\begin{figure}
\includegraphics[width=3.5in]{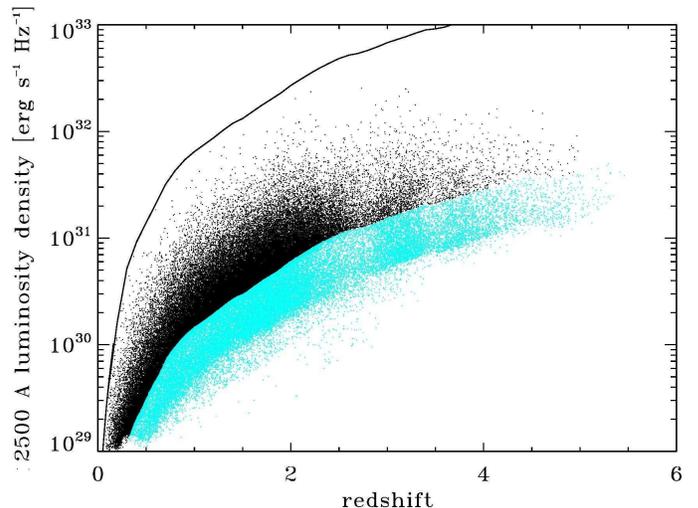}
\caption{The 2500 \AA \, rest frame absolute luminosity density for the SDSS DR7 quasar catalog \citep{SDSSQ}.  Black points are the objects with the $i<19.1$ criterion (63942 objects), while blue points are the objects outside of this subset (41728 objects).  It is seen that the $i<19.1$ subset forms a catalog that has a smoother redshift distribution without a bias toward objects with $z>2$, and with a uniform limiting flux for every redshift \citep[see \S 6 of][]{SDSSQ}.  Also shown is the upper limiting flux corresponding to $i=$15.0.  The 2500 \AA \, luminosity density is obtained from the observed $i$-band magnitude, converting to flux at the integrated center band frequency, and applying the luminosity distance obtained from the redshift with the standard cosmology and the K-corrections provided by R06 which include the continuum and emission line effects.  }
\label{opts}
\end{figure} 

\begin{figure}
\includegraphics[width=3.5in]{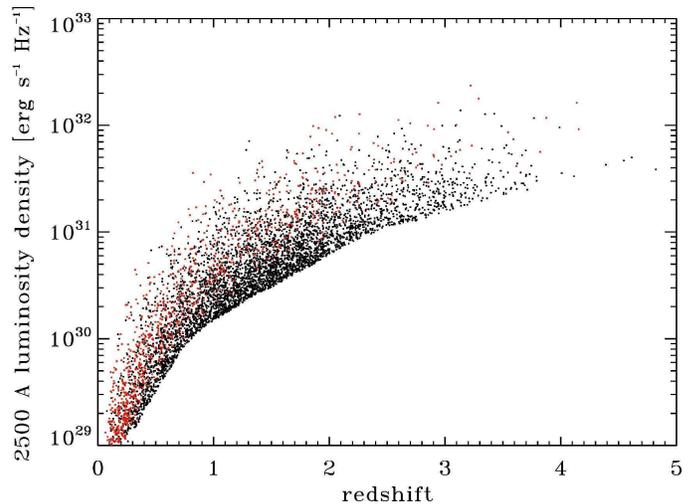}
\caption{Same as Figure \ref{opts} but for the quasars in the canonical SDSS $\times$ FIRST dataset used in this analysis (5445 objects).  The black points are the RL objects while the red points are the RQ.}
\label{optlums}
\end{figure} 

\begin{figure}
\includegraphics[width=3.5in]{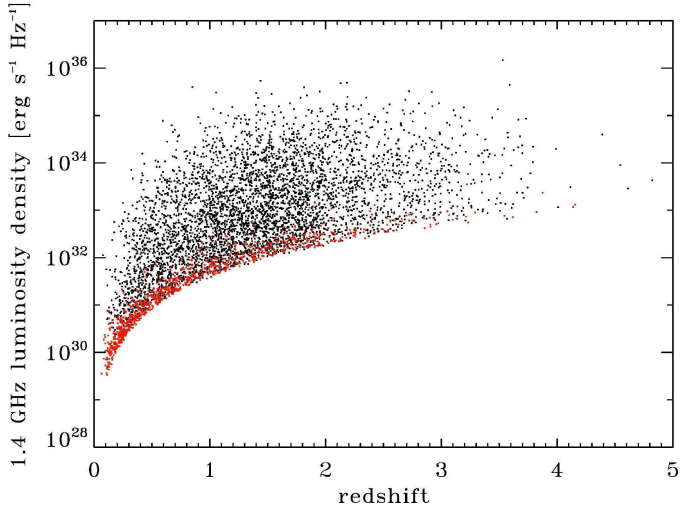}
\caption{The 1.4 GHz rest frame absolute luminosity density for the quasars in the canonical SDSS $\times$ FIRST dataset used in this analysis (5445 objects).  To obtain the 1.4 GHz luminosity density we use the luminosity distance obtained from the redshift and the standard cosmology and the standard K-correction.  We assume a radio spectral index of 0.6.   The black points are the RL objects while the red points are the RQ.}
\label{radlums}
\end{figure} 

\begin{figure}
\includegraphics[width=3.5in]{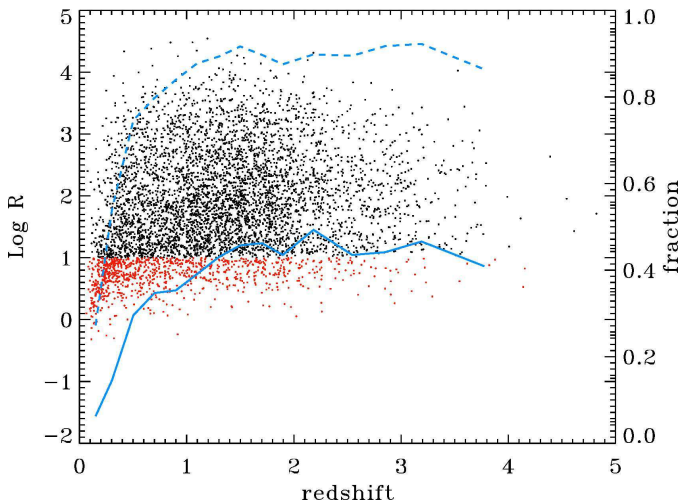}
\caption{The redshift distribution of the ratio $R$ of rest frame absolute luminosities at 5 GHz and 2500
\AA \, for the quasars in the canonical SDSS $\times$ FIRST dataset used in this analysis. The 5 GHz luminosity is obtained from the 1.4 GHz luminosity assuming a radio spectral index of 0.6.  The black points are the RL objects while the red points are the RQ.  Also plotted are the {\it observed} fraction of objects with $R>10$ (dotted curve) and $R>100$ (solid curve). }
\label{sed}
\end{figure} 

\begin{figure}
\includegraphics[width=3.5in]{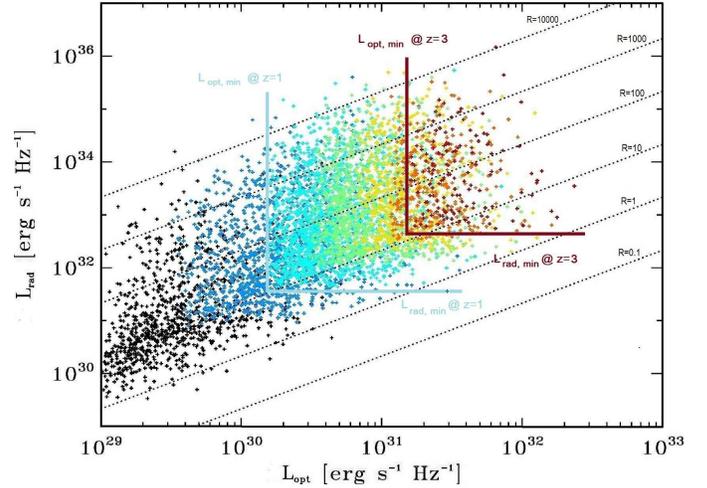}
\caption{The 1.4 GHz rest frame radio luminosity density versus the 2500 \AA \, rest frame optical luminosity density for the quasars in the canonical SDSS $\times$ FIRST dataset used in this analysis.  Colors represent different redshift bins.  Black points are $z \leq 0.5$, dark blue points are $0.5<z \leq 1.0$, light blue points are $1.0<z \leq 1.5$, green points are $1.5<z \leq 2.0$, yellow points are $2.0<z \leq 2.5$, orange points are $2.5<z \leq 3.0$, and red points are $z > 3.0$.  Also shown are lines of constant raw $R$ (defined as the ratio of the 5 GHz radio luminosity to the 2500 \AA \, optical luminosity), and the limiting luminosities for inclusion in the sample at example redshifts of $z=1$ and $z=3$.  The data spans a different range of the de-evolved and bias corrected ratio $R'$ than that for the raw ratio $R$ shown here because of the different redshift evolutions of the radio and optical luminosities (see \S \ref{method}, \ref{Rdist}, and \ref{rldisc}).  The selection effects and their redshift dependence are clearly visible in this figure, while the methods of this work access the true underlying luminosity and density evolutions, luminosity functions, and intrinsic radio loudness distribution. }
\label{LvsL}
\end{figure} 

In order to evaluate the luminosity evolution in both radio and optical, and to separate and compare these effects, we require a dataset that has both radio and optical fluxes to reasonable limits across a broad range of redshifts that contains a significant number of both RL and RQ objects.  The overlap of the FIRST bright quasar radio survey \citep{FIRST1,FIRST2} with the SDSS DR7 quasar catalog \citep[DR7Q;][]{SDSSQ}, can form such a dataset.

The SDSS DR7 quasar catalog  has a limiting $i$ band magnitude of 21 or $f_{m,i}=0.015$ mJy, and contains over 105,000 objects with redshifts ranging from 0.065 to 5.46 over 9380 deg$^2$. We work with a more restrictive $i$-band optical magnitude limit of 19.1 ($f_{m,i}=0.083$ mJy) to form a parent optical catalog that has a smoother redshift distribution with a reduced bias toward objects with $z>$2 and with a uniform limiting flux for every redshift (see \S 6 of DR7Q).  The $i<19.1$ catalog consists of 63,942 objects, with a maximum redshift of 4.98, shown in Figure \ref{opts}.  

In the SDSS DR7 catalog, objects are identified as quasar candidates if either they have the requisite optical colors, {\it or} if they have a radio match within $2''$.  Only 1\% of the quasars in the catalog were selected from the candidate list for spectroscopic followup based on the later criterion.  We also note the presence of an upper limit $i$ band magnitude of 15.0 for inclusion in the catalog.  However this criterion is not completely rigorous, as mentioned in DR7Q.  Further properties of the DR7 quasar catalog, such as black hole masses obtained from line emission measurements, are presented in \citet{Shen11}.  

The FIRST survey, carried out with the Very Large Array (VLA) in B configuration between 1995 and 2004, has a limiting peak pixel 1.4 GHz flux of 1 mJy, and contains 816,000 sources over 9500 deg$^2$.  Different criteria can be used to determine radio and optical matches to form a joint SDSS DR7 QSO $\times$ FIRST catalog.   \citet{Jiang07}, hereafter J07, in their analysis of SDSS quasars, used a matching radius of $5''$ for single radio matches and $30''$ for multiple matches to an optical source.  We construct a fiducial catalog using this radius matching criteria but, unlike J07, we do not include quasars with no radio detection  (see the discussion in \S\,\ref{rldisc}).  

This fiducial catalog contains 5677 objects ranging in redshift from 0.064 to 4.82 and with $\log R$ ranging from $-$0.47 to 4.44.  4327 of the objects in the fiducial catalog are single matches (which J07 calls `Fanaroff-Riley type I sources', FR1s, perhaps incorrectly; see the discussion in \S \ref{rldisc}) and 1350 are multiple matches (which J07 calls `Fanaroff-Riley type II objects', or FR2s for short).  This sample seems to be skewed toward RL as opposed to RQ quasars, with 4543 (80\%) having $R>10$.  It should be noted that J07 used the SDSS Data Release 3, so our realization with the same matching criteria, even with the additional $i<19.1$ magnitude cut and no objects without radio detections, results in more objects (J07 have 2566 objects).

We also construct catalog with a universal $5''$ matching criterion which we feel is more appropriate, as discussed in \S \ref{rldisc}.  In this catalog there are 5445 objects with only 25 multiple radio matches to an optical source.  We will refer to this as the canonical sample, and base our analysis primarily on it, while comparing to results we obtain with the fiducial sample, which as discussed below are quite similar.   In the canonical sample 4466 (82\%) have $R>10$.  The major changes from the fiducial catalog to the canonical one are that some optical objects with multiple radio matches in the fiducial sample are reduced to having only single matches in the canonical sample, with a corresponding reduction in radio flux, while a small number of objects drop out entirely, having had multiple matches within $30''$ but none within $5''$. The fiducial sample and canonical sample are nearly identical in their redshift distribution, and the average radio luminosity differs only slightly, being 10\% lower in the canonical sample.

As shown in \citet{R06}, hereafter R06, e.g. Figure 8 of that work, the SDSS optical quasar sample contains significant biases in the redshift distribution due to emission line effects and the differing flux limits at $z>2$.  We have reduced the later effect by using only the universally flux-limited $i<19.1$ sample, and the former by adopting the full $\varepsilon_{\rm opt}=0.5$ power law continuum {\it plus} emission line K corrections presented in R06 and discussed in \S 5 and Table 4 of that work.  The methods of this work can then account for any bias resulting from emission line effects, as long as they are included in the conversions from luminosity to flux, i.e. in the K corrections.  For the radio data we assume a power law with $\varepsilon_{\rm rad}=0.6$.   Figures \ref{optlums} and \ref{radlums} show the radio and optical luminosities versus redshifts of the quasars in the canonical constructed SDSS x FIRST catalog.  Figure \ref{sed} shows the radio loudness parameter $R$ versus redshift for the canonical dataset, while Figure \ref{LvsL} shows the radio luminosity versus the optical luminosity for different redshift bins.

\section{General remarks on luminosity functions and evolutions}\label{simlumf}

\subsection{Luminosity and density evolution}\label{lde}

The LF gives the number of objects per unit comoving volume $V$ per unit source luminosity, so that the number density is $dN/dV = \int dL_{\rm a} \Psi_{\rm a}(L_a, z)$ and the total number is $N = \int dL_{\rm a} \, \int dz \, (dV/dz) \, \Psi_{\rm a}(L_a, z)$.  To examine luminosity evolution, without loss of generality, we can write a LF in some waveband $a$ as 

\begin{equation}
\Psi_{a}\!(L_a,z) = \rho(z)\,\psi_a\!(L_{a}/g_{a}\!(z) , \eta_a^j)/g_a\!(z),
\label{lumeq}
\end{equation}
where $g_{a}\!(z)$ and $\rho(z)$ describe the luminosity evolution and comoving density evolution with redshift respectively and $\eta_a^j$ stands for parameters that describe the shape (e.g. power law indices and break values) of the $a$ band LF.  In what follows we assume a non-evolving shape for the LF (i.e. $\eta_a^j=$ const, independent of $L$ and $z$), which is a good approximation for determining the global evolutions.  Once the luminosity evolution $g_{a}\!(z)$ is determined using the EP method we can obtain the mono-variate distributions of the independent variables $L'_{\rm a}=L_{\rm a}/g_{\rm a}(z)$ and z, namely the density evolution $\rho(z)$ and ``local'' LF $\psi_{\rm a}$.  The total number of observed objects is then
\begin{equation}
N_{tot} = \int_0^{z_{max}} dz \int_{L_{\rm min}(z)}^\infty { dL_{a} \, \rho(z) \, {dV \over dz} \, { {\psi_a\!\left(L_{a}/g_{a}\!(z)\right)} \over {g_a\!(z)} }  } ,
\label{inteq}
\end{equation}

We consider this form of the LF for luminosities in different bands, allowing for separate (optical and radio) luminosity evolutions.   

In the previous work (QP1) with the White et al. dataset, we assumed a simple power law for the luminosity evolution

\begin{equation}
g_{\rm a}(z)=(1+z)^{k_{\rm a}}.
\label{Levolutionold}
\end{equation}
However with the SDSS dataset which contains objects past $z=4$ and many more objects past $z=3$, we need to use a more complicated parameterization:

\begin{equation}
g_{\rm a}(z)={ {(1+z)^{k_{\rm a}}} \over {1+({{1+z} \over {z_{cr}}})^{k_{\rm a}}} }.
\label{altevolution}
\end{equation} 
With either definition of $g_a(z)$ which gives $g_a(0)=1$, the luminosities $L'_a$ refer to the de-evolved values at $z=0$, hence the name ``local''. 

We have verified the form of equation \ref{altevolution} as a good fit to the evolution by considering the optical evolution in the large parent optical only sample in narrow redshift bins with the appropriate simple form of equation \ref{Levolutionold} and then verifying the form of the cumulative evolution over a larger range.  We have determined the appropriate value for $z_{cr}$ as well as confirmed that the same exponent is appropriate in the numerator and denominator, using the large set of optical only data and the methods of \S \ref{optmeth}.  With the exponents fixed to be the same, the optimal value of $z_{cr}$ is determined to be 3.7$\pm$0.3, by considering the 1$\sigma$ range of the best fit evolution while letting that numerical factor vary.  We would expect that if the truncations and correlations have been properly accounted for by the methods of this work, $g_{\rm opt}\!(z)$ as determined from the simultaneous radio and optical dataset should match that as determined from the parent optical only dataset, which is shown to be the case (see \S \ref{evsec}).

We discuss the determination of the evolution factors $g_a(z)$ with the EP method, which in this parameterization becomes a determination of $k_a$, in \S \ref{evsec}.  The density evolution function $\rho(z)$ is determined by the method shown \S \ref{dev}.  Once these are determined we construct the local (de-evolved) LF $\psi_{\rm a}\!(L_{a}')$, shown in \S \ref{local}.

\subsection{Joint Luminosity Functions}\label{ifcorr}

In general, determination of the evolution of the LF of extragalactic sources for any wavelength band except optical involves a tri-variate distribution because redshift determination requires optical observations which introduces additional observational selection bias and data truncation.  Thus, unless redshifts are known for all sources in a radio survey, we need to determine the combined LF $\Psi\!(L_{\rm opt},L_{\rm rad},z)$ from a tri-variate distribution of $z$ and the fluxes in the optical and radio bands.  If the optical and radio luminosities were statistically independent variables, then this LF would be separable $\Psi\!(L_{\rm opt},L_{\rm rad},z) = \Psi_{\rm opt}(L_{\rm opt}, z) \, \times \, \Psi_{\rm rad}(L_{\rm rad}, z)$ and we would be dealing with two bi-variate distributions.  

However, the luminosities in the different wavebands may be correlated.  The degree and form of the correlation must be determined from the data, and as described below,  the EP method allows us to determine whether any pair of variables are independent or correlated.  Once it is determined that the luminosities are correlated (see \S \ref{roevsec}), the question must be asked whether this luminosity correlation is intrinsic to the population, or induced in the data by flux limits and/or similar luminosity evolutions with redshift.  Determination of which is the case is quite intricate as discussed in Appendix B of QP1, and has not been explored much in the literature. This will be the subject of a future work.  Here we will simply consider both possibilities.  

At one extreme, if the luminosity correlation is intrinsic and not induced, one should seek a coordinate transformation to define a new pair of variables which are independent.  This requires a parametric form for the transformation.  One can define a new luminosity which is a combination of the two, which we call a ``correlation reduced radio luminosity'' $L_{\rm crr}=L_{\rm rad} / F(L_{\rm opt} / L_{\rm fid})$, where the function $F$ describes the correlation between $L_{\rm rad}$ and $L_{\rm opt}$ and $L_{\rm fid}$ is a fiducial luminosity taken here\footnote{This is a convenient choice for $L_{\rm fid}$ as it is lower than the lowest 2500 \AA \, luminosity considered in our sample, but results do not depend on the particular choice of numerical value.} to be $10^{28}$\,erg\,sec$^{-1}$\,Hz$^{-1}$.  For the correlation function we will assume a simple power law 
\begin{equation}
L_{\rm crr} = {{L_{\rm rad}} \over {(L_{\rm opt}/L_{\rm fid})^{\alpha}}}
\label{rcrdef}
\end{equation}
where $\alpha$ is a bulk power law correlation index to be determined from the data.  This is essentially a coordinate rotation in the log-log luminosity space.  As shown in \S \ref{evsec} below, EP also prescribe a method to determine a best fit value for the index $\alpha$ which orthogonalizes the new luminosities.  Given the correlation function we can then transform the data (and its truncation) into the new independent pair of luminosities $(L_{\rm opt}$ and $L_{\rm crr})$.

At the other extreme, if the correlation between the luminosities is entirely induced, then the LFs are separable as described above and the analysis can proceed from there.  As mentioned, we will consider both possibilities here.  It turns out that the results obtained in both cases are very similar, implying that the luminosity-redshift correlations in the analyzed sample are much stronger than any intrinsic luminosity-luminosity correlations (if present).  Further mechanics of the procedure for intrinsically correlated luminosities is discussed in the Appendix.

In this work we are also interested in the distribution of radio loudness parameter $R$, specifically its de-evolved value $R' = L'_{\rm rad}/L'_{\rm opt}$.  In the case of uncorrelated radio and optical luminosities or an induced, non-intrinsic radio-optical luminosity correlation, the local distribution of $R'$, denoted hereafter as $G(R')$, can be constructed from the local optical and radio luminosity functions as in equation \ref{Gtrue}, with the redshift evolution function 

\begin{equation}
g_{\rm R} = { {g_{\rm rad}} \over {g_{\rm opt}} } .
\label{rr}
\end{equation}
In the case of an intrinsic radio-optical luminosity correlation, $G_{\rm R'}$ and the evolution $g_{\rm R}$ can be constructed by equations \ref{localr} and \ref{Rexp} given in the Appendix.

\section{Determination of best fit correlations} \label{evsec}

We now describe results obtained from the use of the procedures described in \S \ref{simlumf} on the data described in \S \ref{datasec}.  Here we first give a brief summary of the algebra involved in the EP method.  We generally follow the procedures described in more detail in QP1.  This method uses the Kendall tau test to determine the best-fit values of parameters describing the correlation functions using  the test statistic 

\begin{equation}
\tau = {{\sum_{j}{(\mathcal{R}_j-\mathcal{E}_j)}} \over {\sqrt{\sum_j{\mathcal{V}_j}}}}
\label{tauen}
\end{equation}
to test the independence of two variables in a dataset, say ($x_j,y_j$) for  $j=1, \dots, n$.  Here $\mathcal{R}_j$ is the dependent variable ($y$) rank of the data point $j$ in a set associated with it.  For untruncated data (i.e. data truncated parallel to the axes) the set associated with point $j$ includes all of the points with a lower (or higher, but not both) independent variable value ($x_k < x_j$).  If the data is truncated one must form the {\it associated set} consisting only of those points of lower (or higher, but not both) independent variabe ($x$) value that would have been observed if they were at the $x$ value of point $j$ given the truncation.  As an example, if we have one sided truncations, then the associated set $A_j \, = \, \{ \, k:x_k \,< \, x_j, \, y^-_k\,<\,y_j \, \}$, where $y^-_k$ is the limiting $y$ value of data point $j$ (see EP for a full discussion of this method). 

If ($x_j,y_j$) were independent then the rank $\mathcal{R}_j$ should be distributed uniformly between 0 and 1 with the expectation value and variance $\mathcal{E}_j=(1/2)(n+1)$ and $\mathcal{V}_j=(1/12)(n^{2}+1)$, respectively, where n is the number of objects in object $j$'s associated set.  Independence is rejected at the $m \, \sigma$ level if $\vert \, \tau \, \vert > m$.  To find the best fit correlation the $y$ data are then adjusted by defining $y'_j=y_j/F(x_j)$  and the rank test is repeated, with different values of parameters of the function $F$.

\subsection{Optical-Only Dataset Luminosity-Redshift Correlation} \label{optmeth}

With the optical only dataset (here we use the $i<19.1$ set described in \S \ref{datasec}), determination of the luminosity-redshift correlation function $g_{\rm opt}\!(z)$ by equation \ref{altevolution} reduces to determination of the value of the index $k_{\rm opt}$.  As evident from Figure \ref{opts} the $L_{\rm opt}-z$ data are heavily truncated due to the flux limits of SDSS.   The associated set for object $j$ includes only those objects that are sufficiently luminous to exceed the optical flux minimum for inclusion in the survey if they were located at the redshift of the object in question, ie $L_k \geq L_{min}(z_j)$.  As the values of $k_{\rm opt}$ are adjusted and $L_{\rm opt}$ is scaled by $g_{\rm opt}\!(k,z)$, the luminosity cutoff limits for a given redshift are also scaled by $g_{\rm opt}\!(k,z)$.  

\begin{figure}
\includegraphics[width=3.5in]{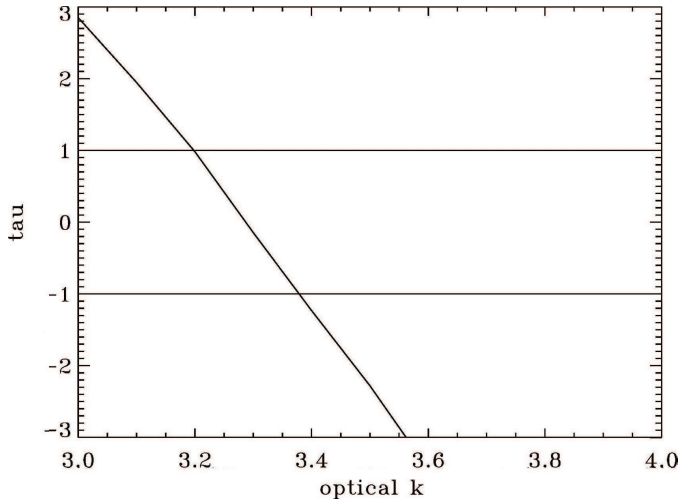}
\caption{The value of the $\tau$ statistic as given by equation \ref{tauen} as a function of $k_{\rm opt}$ for the form of the optical luminosity evolution given by equation \ref{altevolution}, for the 63,942 quasars in the $i<19.1$ optical only dataset.  The 1 $\sigma$ range for the best fit value of $\alpha$ is where $\vert \, \tau \, \vert \leq 1$.   } 
\label{optonly}
\end{figure}

Figure \ref{optonly} shows the value of the test statistic $\tau$ for a range of values of $k_{\rm opt}$ given the parameterization of equation \ref{altevolution}.  It is seen that the best fit value of $k_{\rm opt}$ is 3.3 with a 1$\sigma$ range of 0.1.  This indicates that quasars have strong optical luminosity evolution, i.e. the LF shifts to higher luminosities at larger redshifts.

We have performed an optimization procedure to determine the best-fit value of $z_{cr}$, the numerical constant in the denominator of equation \ref{altevolution}, as well as whether  the same exponent is appropriate for the numerator and denominator.   This was done by varying these parameters and determining the best-fit range of values in the same way as described above.  We find $z_{\rm cr}$ = 3.7$\pm$0.3.

In this analysis we have ignored the upper optical flux limit of SDSS quasars corresponding roughly to a $i$ band magnitude of 15.0 as discussed in \S \ref{datasec}.  The reason for this is that data truncations are only consequential in this analysis if the truncation is actually denying the sample data points that would otherwise be there.  As can be seen in Figure \ref{opts}, there are very few objects approaching this upper truncation limit, indicating that the truncation is not consequential, i.e. it does not appreciably alter the sample from the underlying population.  

\subsection{Radio-Optical Luminosity Correlation} \label{roevsec}

\begin{figure}
\includegraphics[width=3.5in]{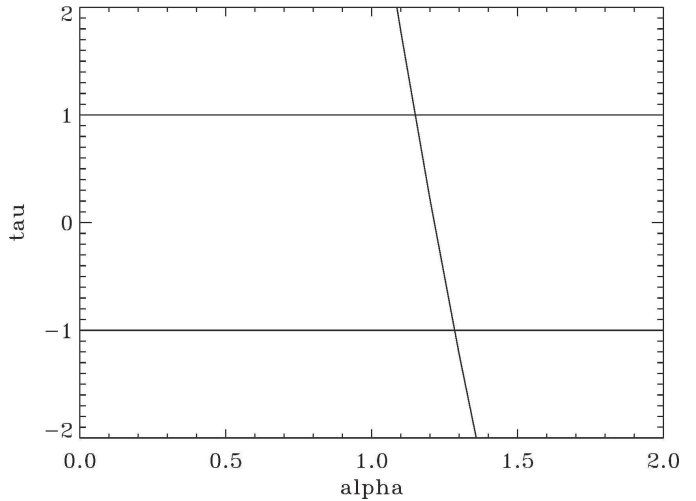}
\caption{The value of the $\tau$ statistic as given by equation \ref{tauen} as a function of $\alpha$ for the relation $L_{\rm rad} \propto (L_{\rm opt})^{\alpha}$, where $L_{\rm opt}$ and $L_{\rm rad}$ are the optical and radio luminosities, respectively, for the quasars in the canonical combined dataset.  The 1$\sigma$ range for the best fit value of $\alpha$ is where $\vert \, \tau \, \vert \leq 1$.  It is seen that the observed optical and radio luminosities are strongly positively correlated, with a linear relation, although this may not be the intrinsic correlation, as discussed in \S \ref{roevsec}, \S \ref{ifcorr}, and the Appendix. } 
\label{taus}
\end{figure}

Turning to the combined radio and optical dataset, we will determine the correlation between the radio and optical luminosities.  The radio and optical luminosities are obtained from radio and optical fluxes from a two-flux limited sample so that the data points in the two dimensional flux space are truncated parallel to the axes which we consider to be untruncated.  Since the two luminosities have essentially the same relationship with their respective fluxes, except for a minor difference in the K-correction terms, we can consider the luminosity data to also be untruncated in the $L_{\rm opt}$-$L_{\rm rad}$ space.  In that case the determination of the associated set is trivial and one is dealing with the standard Kendall tau test.  Assuming the correlation function between the luminosities $F(x)=x^{\alpha}$ we calculate the test statistic $\tau$ as a function of  $\alpha$.  Figure \ref{taus} shows the absolute value of the test statistic $\tau$ vs $\alpha$, from which we get the best fit value of $\alpha=1.2$ with one $\sigma$ range $\pm 0.1$.  The result similar whether the canonical or fiducial combined dataset is used.  As expected $\alpha$ is near unity, and this result is compatible with that obtained in QP1 ($\alpha=1.3 \pm 0.2$).  

As discussed in \S \ref{ifcorr}, this correlation may be inherent in the population or may be an artifact of the flux limits and wide range of redshifts.  The general question of determining whether an observed luminosity correlation is intrinsic or induced will be explored in a future work.  For this analysis going forward, we will consider both possibilities.  We see that it does not make a significant difference for the major conclusions of this work.

\subsection{Joint Dataset Luminosity-Redshift Correlations} \label{method}

The basic method for determining simultaneously the best fit $k_{\rm opt}$ and $k_{\rm rad}$, given the evolution forms in equation \ref{altevolution}, is the same as described in \S \ref{optmeth} but in this case the procedure is more complicated because we now are dealing with a three dimensional distribution  ($L_{\rm rad}, L_{\rm opt}, z$) and two correlation functions ($g_{\rm rad}\!(z)$ and $g_{\rm opt}\!(z)$).

Since we have two criteria for truncation, the associated set for each object includes only those objects that are sufficiently luminous in both bands to exceed {\it both} flux minima for inclusion in the survey if they were located at the redshift of the object in question.  Consequently, we have a two dimensional minimization problem, because both the optical and radio evolution factors, $g_{\rm opt}\!(z)$ and $g_{\rm rad}\!(z)$, come into play, as the luminosity cutoff limits for a given redshift are also adjusted by powers of $k_{\rm opt}$ and $k_{\rm rad}$.  

\begin{figure}
\includegraphics[width=3.5in]{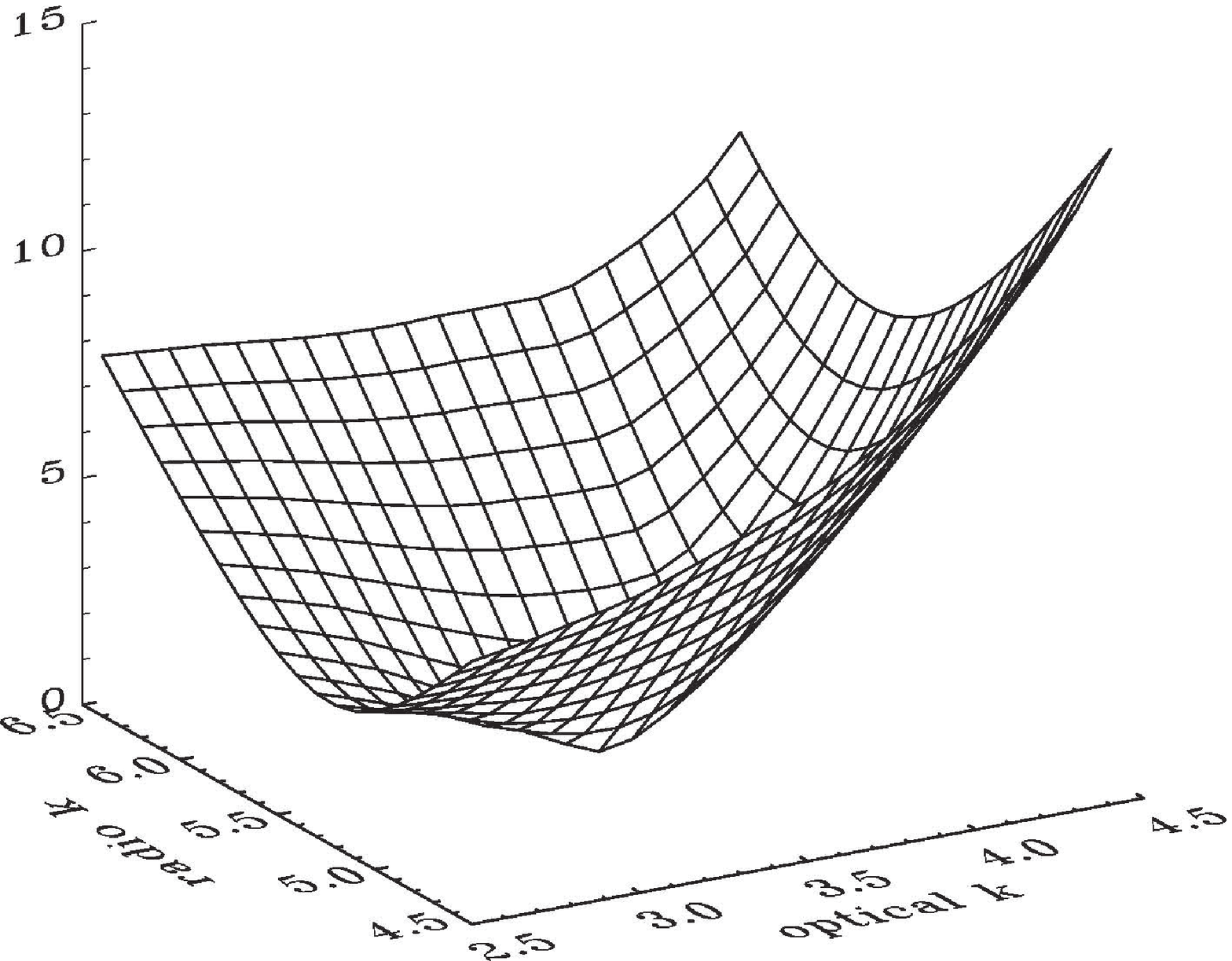}
\caption{Surface plot of the value of $\tau_{\rm comb}$ for the canonical combined radio-optical dataset as a whole showing the location of the minimum region where the favored values of $k_{\rm opt}$ and $k_{\rm rad}$ lie.}
\label{tausss}
\end{figure}

\begin{figure}
\includegraphics[width=3.5in]{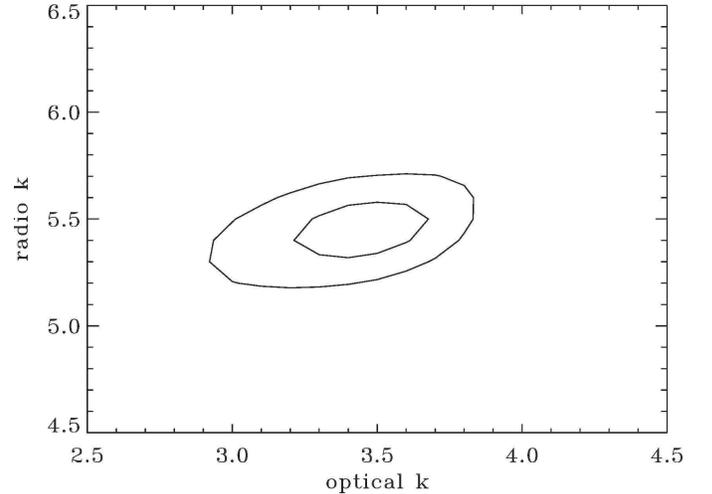}
\caption{The $1\sigma$ and $2 \sigma$ contours for the simultaneous best fit values of $k_{\rm opt}$ and $k_{\rm rad}$ for the canonical combined dataset, for the forms of the luminosity evolutions  given by equation \ref{altevolution}.   }
\label{alphas}
\end{figure}

We form a test statistic $\tau_{\rm comb} = \sqrt{\tau_{\rm opt}^2 + \tau_{\rm rad}^2} $ where $\tau_{\rm opt}$ and $\tau_{\rm rad}$ are those evaluated considering the objects' optical and radio luminosities, respectively.  The favored values of $k_{\rm opt}$ and $k_{\rm rad}$ are those that simultaneously give the lowest $\tau_{\rm comb}$ and, again, we take the $1 \sigma$ limits as those in which $\tau_{\rm comb} \, < 1$.  For visualization, Figure \ref{tausss} shows a surface plot of $\tau_{\rm comb}$.  Figure \ref{alphas} shows the best fit values of $k_{\rm opt}$ and $k_{\rm rad}$ taking the 1 and 2 $\sigma$ contours, for the canonical radio-optical dataset.   We have verified this method with a simulated Monte Carlo dataset as discussed in QP1.

We see that positive evolution in both radio and optical wavebands is favored.  The minimum value of $\tau_{\rm comb}$ favors an optical evolution of $k_{\rm opt}$ = 3.5 and a radio evolution of $k_{\rm rad}$ = 5.5, with a very small uncertainty at the $1 \sigma$ level.  The fact that  $k_{\rm opt}$ as determined here from the combined dataset is equal to that determined from the much larger optical only dataset (see \S \ref{optmeth} and Figure \ref{optonly}) indicates that the truncations have been properly handled.  

In the above analysis we have assumed  sharp truncation boundaries and that the data is complete above the boundaries.  As discussed in \S \ref{testass} this may not be the case for the FIRST radio data.  If we restrict the sample to only data with $f_{\rm rad} > 2$ mJy the favored optical evolution range for the canonical dataset lowers slightly to $k_{\rm opt}=3.0 \pm 0.5$ and the favored radio evolution range lowers slightly to $k_{\rm rad}=5.0 \pm 0.25$.  These overlap with the results considering the entire combined canonical sample at the 1 $\sigma$ level.

These results indicate that quasars have undergone a significantly greater radio evolution relative to optical evolution, in general agreement with the result in QP1, although the best-fit differential evolution was even more dramatic there, given the form of $g(z)$ used.  

The results for the best fit $k_{\rm opt}$ and $k_{\rm rad}$ are identical if the different fiducial radio-optical dataset matching radius criteria are used, indicating that the addition of extra radio flux and a few additional sources do not matter for this determination.  If we consider that the radio-optical luminosity correlation is inherent in the data, then the orthogonal luminosities are $L_{\rm opt}$ and $L_{\rm crr}$ (see \S \ref{ifcorr}) and we can determine the best fit evolutions $g_{\rm crr}\!(z)$ and $g_{\rm opt}\!(z)$.  These results favor $k_{\rm opt}=3.5$ and $k_{\rm crr}=2$.   In this case the best fit radio evolution can be recovered by equation \ref{gradform} in the Appendix and it is $k_{\rm rad}=5.5$, in perfect agreement with the results obtained from considering $k_{\rm opt}$ and $k_{\rm rad}$ as orthogonal, indicating the robustness of the result.

\section{Density evolution} \label{dev}

Next we determine the density evolution $\rho(z)$.  One can define the cumulative density function 

\begin{equation}
\sigma(z) = \int_0^z { {{dV} \over {dz}} \, \rho\!(z) \, dz}
\end{equation}
which, following \citet{P92} based on \citet{L-B71}, can be calculated by

\begin{equation}
\sigma(z) = \prod_{j}{(1 + {1 \over m(j)})}
\label{sigmaeqn}
\end{equation}
where $j$ runs over all objects with a redshift lower than or equal to $z$, and $m(j)$ is the number of objects with a redshift lower than the redshift of object $j$ {\it which are in object j's associated set}.  In this case, the associated set is again those objects with sufficient optical and radio luminosity that they would be seen if they were at object $j$'s redshift.  The use of only the associated set for each object removes the biases introduced by the data truncation.  Then the density evolution $\rho(z)$ is 

\begin{equation}
\rho\!(z) = {d \sigma\!(z) \over dz} \times {1 \over dV/dz}
\label{rhoeqn}
\end{equation}

\begin{figure}
\includegraphics[width=3.5in]{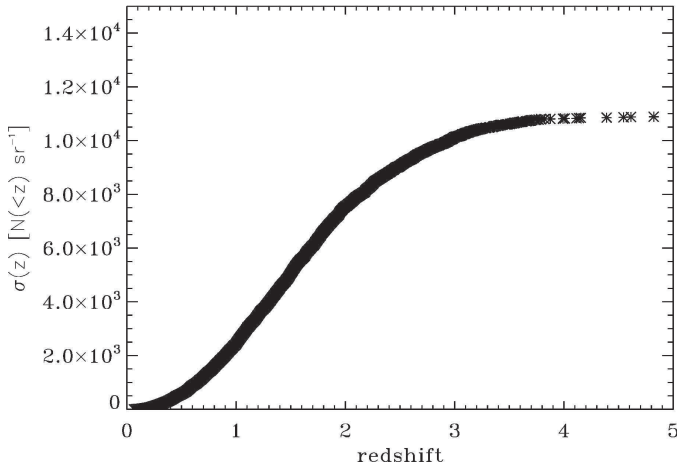}
\caption{The cumulative density function $\sigma(z)$ vs. redshift for the quasars in the canonical dataset.  The normalization of $\sigma(z)$ is determined as described in \S \ref{dev}.  A polynomial fit to $\sigma\!(z)$ is used to determine $\rho\!(z)$ by equation \ref{rhoeqn}.  }
\label{sigma}
\end{figure}

\begin{figure}
\includegraphics[width=3.5in]{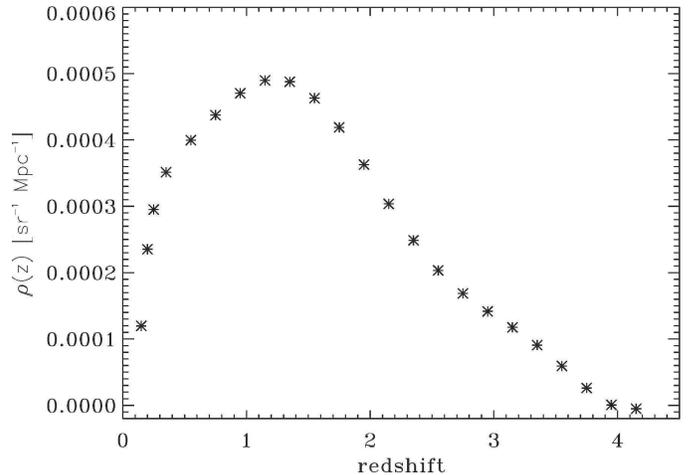}
\caption{The density evolution $\rho(z)$ vs. redshift for the for the quasars in the canonical dataset.  $\rho(z)$ is defined such that $\sigma(z)=\int_0^{\infty} \rho(z) \, dV/dz \, dz$.  The normalization of $\rho(z)$ is determined as described in \S \ref{dev}, and polynomial fits of $\rho(z)$ to $z$ are given there. }
\label{rholog}
\end{figure}

However, to determine the density evolution, the previously determined luminosity evolution must be taken out.  Thus, the objects' optical and radio luminosities, as well as the optical and radio luminosity limits for inclusion in the associated set for given redshifts in the calculation of $\sigma$ by equation \ref{sigmaeqn}, are scaled by taking out factors of $g_{\rm opt}\!(z)$ and $g_{\rm rad}\!(z)$,  determined as above.  Figures \ref{sigma} and \ref{rholog} show the cumulative and differential density evolutions, respectively.\footnote{We note a notional difference with QP1 where we plotted $\rho(z)=d\sigma / dz$ rather than $\rho$ as defined in equation \ref{rhoeqn} here with a factor of $dV/dz$ taken out.}  The number density of quasars seems to peak at between redshifts 1 and 1.5, earlier than generally thought for the most luminous quasars \citep[e.g.][]{Shaver96}, and earlier than that found in R06, but more similar to the peak found for less luminous quasars by \citet{Hopkins07}, and in agreement with \citet{MP99}.  This is slightly earlier than we found in QP1.

The normalization of $\rho(z)$ is determined by equation \ref{inteq}, with the customary choice of $\int_{L_{\rm min}'}^{\infty} {\psi(L') \, dL'}=1$.  As stated in R06, the main sources of bias in the redshift distribution of the SDSS quasar sample are 1) the differing magnitude limits for $z>2$, 2) the effects of emission lines on $i$ band flux at different redshifts, and, at a somewhat less important level, 3) the inclusion of extended sources at the lowest redshifts ($z \sim 0.3$).  As discussed in \S \ref{datasec}, we have, following R06, dealt with issue 1 by restricting the sample to a universal $i<19.1$ magnitude limit, and issue 2 by adopting the R06 K corrections which include the effect of emission lines as well as the continuum spectrum.  Since we do not see any hint of a peak or deviation in $\rho(z)$ or $\sigma(z)$ at $z \sim 0.3$, we do not deem issue 3 to significantly effect our conclusions.  The density evolution $\rho(z)$ can be fit with a broken polynomial in $z$, with coefficients shown in Table 1.  

\begin{figure}
\includegraphics[width=3.5in]{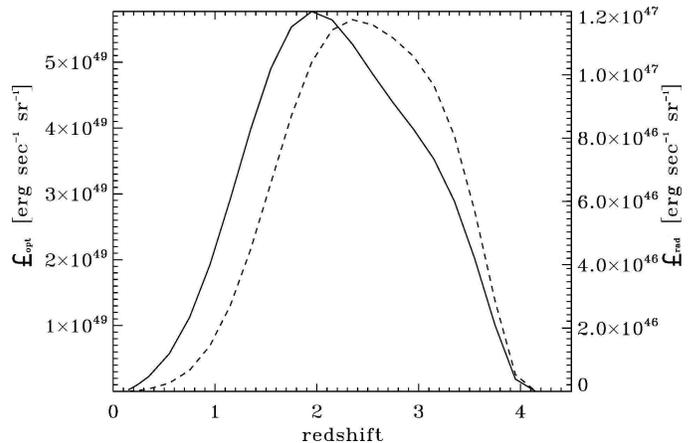}
\caption{The optical (solid curve) and radio (dashed curve) luminosity densities $\pounds_{\rm opt}(z)$ and $\pounds_{\rm rad}(z)$, discussed in \S \ref{dev} and equation \ref{fancyLeq}.  Note the different scales. }
\label{fancyL}
\end{figure}

Knowing both the luminosity evolutions $g_{\rm a}\!(z)$ and the density evolution $\rho(z)$, one can form the luminosity density functions $\pounds_{\rm a}(z)$ which are the total rate of production of energy of quasars as a function of redshift, by

\begin{equation}
\pounds_{\rm a}(z) = \int dL_{\rm a} \,  {L_{\rm a} \, \rho(z) \, {dV \over dz}} \, dz \, = \, \langle L'_{\rm a} \rangle \, g_a(z) \, \rho(z) \, {dV \over dz} .
\label{fancyLeq}
\end{equation}
We show both the 2500 \AA \, and optical and 1.4 GHz radio luminosity density functions in Figure \ref{fancyL}.  As expected, because of the greater luminosity evolution in the radio band, the radio luminosity density peaks earlier in redshift than the optical luminosity density.

\begin{deluxetable}{lcccccccc}\label{tizzable}
\tabletypesize{\scriptsize}
\tablecaption{Coefficients for polynomial fit\tablenotemark{a} to density evolution $\rho(z)$ vs. $z$  } 
\tablecolumns{3}
\startdata
  & $z \leq 1.3$ & $z \geq 1.3$ \\
\hline
c & -0.000581965 & 0.00487286 \\
$a_1$ & +0.00752870  & $-$0.0145420  \\
$a_2$ & $-$0.0233918  & +0.0196557  \\
$a_3$ & +0.0355617  & $-$0.0138312  \\
$a_4$ & $-$0.0257837  & +0.00544187  \\
$a_5$ & +0.00714230  & $-$0.0012073  \\
$a_6$ & 0  & +0.000140803  \\
$a_7$ & 0  &  $-$6.68658e-006  \\
\enddata
\tablenotetext{a}{Polynomial fits are of the form $\rho(z) = c_1 \, + \, a_1 \, z \, + \, a_2 \, z^2 \, + \, a_3 \, z^3 \, + \, a_4 \, z^4 \, + \, a_5 \, z^5 \, + \, a_6 \, z^6 \, + \, a_7 \, z^7  $.}
\end{deluxetable}

\section{Local luminosity functions} \label{local}

\subsection{General Considerations}

In a parallel procedure we can use the local (redshift evolution taken out, or 'de-evolved') luminosity distributions (and de-evolved luminosity thresholds) to determine  the `local' LFs $\psi_{a}\!({L_a}')$, where again the subscript $a$ denotes the waveband, and the prime indicates that the luminosity evolution has been taken out.  We first obtain the cumulative LF 

\begin{equation}
\Phi_a\!(L_{a}') = \int_{L_{a}'}^{\infty} {\psi_a\!(L_{a}'') \, dL_{a}''}
\end{equation}
which, following \citet{P92}, $\Phi_a\!(L_{a}')$, can be calculated by

\begin{equation}
\Phi_a\!(L_{a}') = \prod_{k}{(1 + {1 \over n(k)})}
\label{phieq}
\end{equation}
where $k$ runs over all objects with a luminosity greater than or equal to $L_a$, and $n(k)$ is the number of objects with a luminosity higher than the luminosity of object $k$ which are in object $k$'s associated set, determined in the same manner as above. 
The luminosity function $\psi_a\!(L_{a}')$ is 

\begin{equation}
\psi_a\!(L_{a}') = - {d \Phi_a\!(L_{a}') \over dL_{a}'}
\label{psieqn}
\end{equation}

In \S \ref{evsec} we determined the luminosity evolution for the optical luminosity $L_{\rm opt}$ and the radio luminosity $L_{\rm rad}$.   We can form the local optical $\psi_{\rm opt}\!(L_{\rm opt}')$ and radio $\psi_{\rm rad}\!(L_{\rm rad}')$ LFs straightforwardly, by taking the evolutions out.  As before, the objects' luminosities, as well as the luminosity limits for inclusion in the associated set for given redshifts, are scaled by taking out factors of $g_{\rm rad}\!(z)$ and $g_{\rm opt}\!(z)$, with $k_{\rm rad}$ and $k_{\rm opt}$ determined in \S \ref{evsec}.  We use the notation $L \rightarrow L' \equiv L/g(z) $.  For the local luminosity functions, we use the customary normalization $\int_{L_{\rm min}'}^{\infty} {\psi(L') \, dL'}=1$.  This normalization may be biased by around 8\% due to quasar variability as discussed in \S \ref{testass}.

\subsection{Local optical luminosity function}\label{locopt}

\begin{figure}
\includegraphics[width=3.5in]{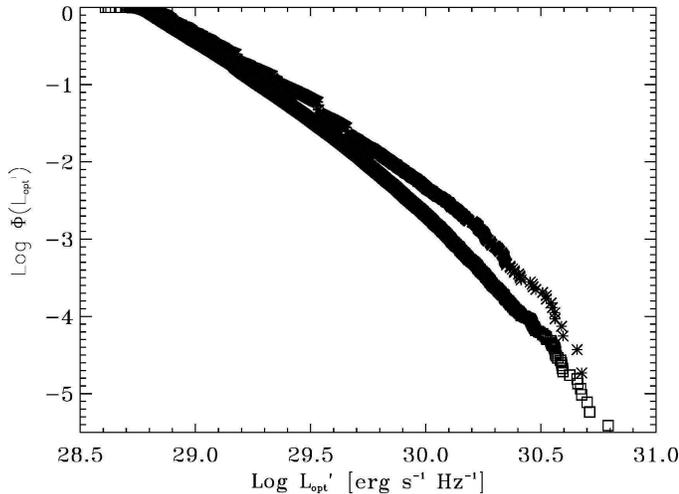}
\caption{The cumulative local optical luminosity function $\Phi_{\rm opt}\!(L_{\rm opt}')$ for the quasars in the canonical combined dataset (stars) and for the parent optical only dataset (squares).  A piecewise quadratic fit to $\Phi\!(L_{\rm opt}')$ is used to determine $\psi_{\rm opt}\!(L_{\rm opt}')$ by equation \ref{psieqn}. The normalization of the optical luminosity function may be biased by around 8\% (see \S \ref{testass}.) }
\label{phiopt}
\end{figure}

\begin{figure}
\includegraphics[width=3.5in]{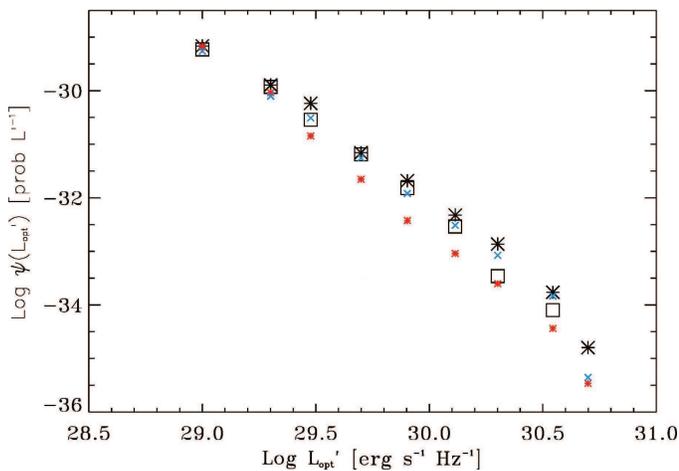}
\caption{The local optical luminosity function $\psi_{\rm opt}\!(L_{\rm opt}')$.  The squares show results for the parent optical only dataset, while the larger black stars, smaller red stars, and smaller blue stars show the results for the entire canonical combined radio-optical dataset, and that dataset with imposed radio flux limits of 2 mJy and 4 mJy, respectively.   The normalization of the local luminosity functions is described in \S \ref{local}.  }
\label{psiopt}
\end{figure}

Figures \ref{phiopt} and \ref{psiopt} show the local cumulative $\Phi_{\rm opt}\!(L_{\rm opt}')$ and differential $\psi_{\rm opt}\!(L_{\rm opt}')$ local optical LFs of the quasars in the canonical combined dataset, and that of the optical only dataset.  

The optical LF shows possible evidence of a break at $\sim 10^{30}$\,erg\,s$^{-1}$\,Hz$^{-1}$, which was present already in data analyzed in \citet{P73} and was also seen in QP1.  With the optical only dataset,  fitting a broken power law yields values $-2.8\pm 0.2$ and $-4.1\pm 0.4$ below and above the break, respectively.  With the canonical combined radio-optical dataset, fitting a broken power law yields values  $-2.8\pm 0.3$ and $-3.8\pm 0.5$.  If we allow for the possibility of additional uncertainty resulting from the consideration of possible radio incompleteness at faint fluxes (see discussion in \S \ref{testass}), the range on the power law below the break changes to $-2.1\pm 0.1$, and above the break the results are not affected.   

As the optical LF has been studied extensively in various AGN surveys, we can compare the slope of $\psi_{\rm opt}\!(L_{\rm opt}')$ obtained here to values reported in the literature.  In QP1, we found power law values of $-2.0\pm0.2$ and $-3.2\pm0.2$ below and above the break, respectively. \citet{Boyle00}, using the 2dF optical dataset (but with no radio overlap criteria) use a customary broken power law form for the LF, with values ranging from $-$1.39 to $-$3.95 for different realizations, showing reasonable agreement.\footnote{It should be noted that they parameterize evolution differently and work in absolute magnitudes rather than luminosities, however the slopes of their fits to the LF as they parameterize it are applicable, as can be seen in their section 3.2.2.}

\subsection{Local radio luminosity function}\label{localradio}

\begin{figure}
\includegraphics[width=3.5in]{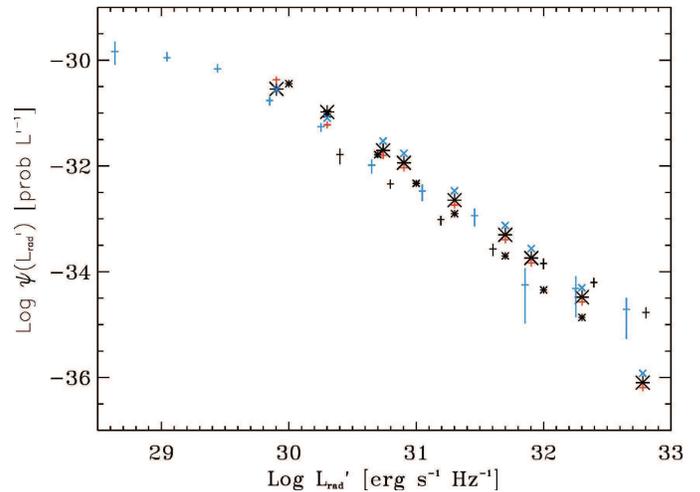}
\caption{The local radio luminosity function $\psi_{\rm rad}\!(L_{\rm rad}')$ for the quasars in the canonical combined radio-optical dataset.  The larger black stars show the results from considering the entire canonical combined dataset and the radio-optical luminosity correlation to be entirely induced, while the small stars show the results from the entire canonical combined dataset considering the correlation to be entirely intrinsic, calculated with equation \ref{localrad} in the Appendix.  The smaller red and blue stars show the results from considering an induced correlation with the canonical combined dataset and imposed radio flux limits of 2 mJy and 4 mJy respectively.  It is seen that the local radio luminosity function is not sensitive to the assumption of whether the radio-optical luminosity correlation is intrinsic or induced (see \S \ref{ifcorr} for a discussion).  The normalization of the local luminosity functions is described in \S \ref{local}. We also overplot the local radio luminosity functions as determined by \citet{Kimball11} with an SDSS $\times$ NVSS sample (black crosses) and by a deep EVLA sample (dark blue crosses), which have been properly scaled for overplotting here as discussed in \S \ref{localradio}. }
\label{psirad}
\end{figure}

Figure \ref{psirad} shows the local radio LF $\psi_{\rm rad}\!(L_{\rm rad}')$.  It is seen that the local radio LF contains a possible break around $2 \times 10^{31}$\,erg\,sec$^{-1}$\,Hz$^{-1}$, with a power law slope of $-1.5\pm0.1$ below the break and $-2.6\pm0.1$ above the break.  The range for the power law above the break is increased slightly to $-2.5\pm 0.3$ if the effects of possible radio incompleteness are included, as in \S \ref{testass}, while below the break it is unchanged, indicating this is a small effect.  It is interesting to note that the value of the radio loudness parameter $R$ corresponding to the break in the radio and optical luminosity functions is very close to the critical value $R \sim 10$ widely discussed in the context of the RL/RQ dichotomy.

We can also construct the local radio LF from $\psi_{\rm opt}\!(L_{\rm opt}')$ and $\psi_{\rm crr}\!(L_{\rm crr}')$  with equation \ref{localrad} of the Appendix, under the assumption that the radio and optical luminosity correlation is intrinsic.  This is shown by the small black points in Figure \ref{psirad}.  It is seen that $\psi_{\rm rad}\!(L_{\rm rad}')$ constructed in this way is nearly identical to that determined by considering the radio and optical luminosity correlation is induced, indicating that the radio luminosity function result is robust to this consideration.

In QP1 for the local radio luminosity function we found a break at $10^{31}$\,erg\,sec$^{-1}$\,Hz$^{-1}$, with a power law slope of $-1.7\pm0.1$ below the break and $-2.4\pm0.1$ above it, in excellent agreement with the results here.  The slope above the break seen here is similar to earlier results of \citet{Schmidt72} and \citet{P73} which probed only high luminosities.  A more complete comparison can be made with \citet{MS07}, who form radio LFs of local sources in the Second Incremental Data Release of the 6 degree Field Galaxy Survey (6dFGS) radio catalog.  For the sources they identify as AGN, they find a break at $3.1 \times 10^{31}$\,erg\,sec$^{-1}$\,Hz$^{-1}$, with slopes of $-2.27\pm0.18$ and $-1.49\pm0.04$ above and below the break, in good agreement with our results.  

In Figure \ref{psirad} we also show the radio luminosity functions for the same range of $L_{rad}'$, as determined by \citet{Kimball11} for two samples, one for a sample of SDSS optical and NRAO Very Large Array Sky Survey (NVSS) radio data, and another for deep radio observations with the Extended Very Large Array in which almost every SDSS quasar in a narrow redshift range in a field was detected in radio.  The results presented there are the luminosity function only for sources in the redshift range $0.2<z<0.3$, which we have scaled by means of equation \ref{altevolution} and converted from 6 GHz to 1.4 GHz with a radio spectral index of 0.6 to obtain the local luminosity function at 1.4 GHz for direct comparison with our results.  It is seen that our luminosity functions agree on the faint end of the range considered here but potentially disagree only on the very bright end.

\section{Distribution of radio loudness ratios}  \label{Rdist}

\begin{figure}
\includegraphics[width=3.5in]{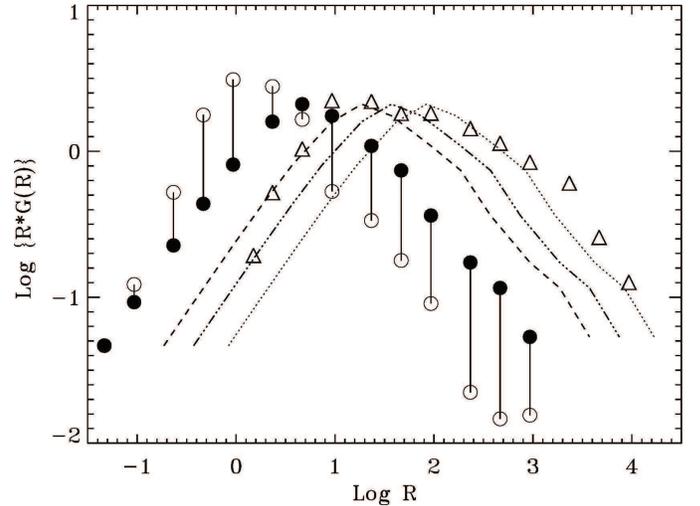}
\caption{The local distribution $G_{R}\!(R)$ in the 5 GHz radio to 2500 \AA \, optical luminosity ratio $R$ for the quasars in the dataset, plotted as $R \times G_{R}\!(R)$.  The circles are from $G_{R'}\!(R')$  as determined by the method of equation \ref{localr}, taking account of the truncations and correlations in the luminosity evolutions, while the triangles result from forming a distribution with a naive use of the objects' raw ratio.  The error bars represent the extremal possibilities of the radio-optical luminosity correlation being entirely induced (open circles) and entirely intrinsic (closed circles).  The normalization is arbitrary.  Also shown is the proper radio loudness distribution $G_R\!(R,z)$ at redshifts $z=1$ (dashed line) and $z=3$ (dash-dot line), and $z=5$ (dotted line) evolved according to the form of equation \ref{Rexp}.  A comparison with data from the literature is shown in Figure \ref{psir2}. }
\label{psir}
\end{figure}

As stated in the introduction, naively one may expect that because the ratio $R$ is independent of cosmological model and nearly independent of redshift, the raw observed distribution of $R$ would provide a good representation of the true distribution of this ratio.  In Appendix A of QP1 we show that this assumption would likely be wrong.  In Figure \ref{psir} we show this raw distribution of $R$ by the triangles, arrived at by using the raw values of $R$ from the data and forming a distribution in the manner of equations \ref{phieq} and \ref{psieqn} with no data truncations accounted for.  As discussed in \S \ref{simlumf}, we can reconstruct the local distribution of $G_{R'}\!(R')$, as in equation \ref{Gtrue}, which provides for a more proper accounting of the biases and truncations and is an estimate of the true, intrinsic distribution.  The results of this calculation are also shown in Figure \ref{psir} by the filled circles.  We include a range of error for the extremal possibilities that the correlation between the radio and optical luminosities is entirely induced or entirely intrinsic (open circles).  In the later case, $G_{R'}\!(R')$ is given by equation \ref{localr} of the Appendix.  

The distribution $G_{R'}\!(R')$ calculated from the data taking into account the evolutions and truncations is clearly different than the raw distribution, and shows no evidence of bi-modality in the range of $R$ considered.  The raw distribution is seen to be weighted toward much higher values of $R$.  Using the naive, raw distribution instead of the reconstructed intrinsic one would result in estimates of quasar properties that would be significantly and systematically biased.

We also know that the best fit redshift evolution of the ratio, given equation \ref{Rexp}.  The change in the distribution of $R$ with increasing redshift is also shown in Figure \ref{psir}.\footnote{Note that we have not included the density evolution which will shift the curves vertically but not change their shape.}  Another way to look at this is that we have found that the radio luminosity evolves at a different rate than the optical luminosity, with the consequence that their ratio is a function of redshift.   The radio loudness of the population increases by a factor of 3 by redshift 1, a factor of 8 by redshift 3, and a factor of 11 by redshift 5.  This is a less dramatic evolution of the ratio than we found in QP1, where we used a simpler parameterization (see \S \ref{lde}), although still quite significant.  The general trend toward increasing $R$ with redshift can be validated in a simple way by examining the median value of $R$ vs. redshift in bins of redshift which indeed shows a steady increase.

This differential evolution is in disagreement with the result presented by J07 who show a decrease in fraction of RL sources with increasing redshift, which could be the case if the radio luminosity were to evolve more slowly than the optical luminosity. They however do not determine individual evolutions or LFs.  On the other hand, \citet{Miller90}  have noted that the fraction of RL quasars may increase with redshift, which they attribute to a difference in the evolutions of the two populations (RL and RQ).  \citet{Donoso09} compute radio and optical LFs at different redshifts and reach the same conclusion.  \citet{Cira06} also find that the radio loud fraction may modestly increase at high redshift.  Although not directly comparable, \citet{LaFranca10} show a similar evolution for $R_x$, the ratio of radio to X-ray luminosity, as we show here for $R$.  As discussed in \S \ref{rldisc}, we believe that our differences with J07 stem from our inclusion of only radio-detected quasars in the joint radio-optical sample.  

We note that our results favor one population in the range of $R$ considered here, in the sense that the distribution of $G(R)$, recovered from considering the data truncations inherent in the survey and correlations between the luminosities, is continuous and without any bi-modality.  This is in agreement with the conclusion we reached in QP1.  One may hypothesize about a bi-modality that is not centered on the commonly assumed value of $R=10$ or above that value, but at a lower value of $R$ below those values probed in this analysis.  Our results disfavor that as well.  As seen in Figures \ref{psiopt} and \ref{psirad}, the lack of a significant change in the shapes of the computed local optical or radio LFs when the radio flux limits for inclusion in the analysis are changed, or the parent optical only dataset is considered, argues against the presence of an additional population of quasars which would become more prominent below the lowest values of $R$ considered here.  The results of \citet{Kimball11}, which probe even lower radio fluxes, and extend the radio luminosity function to lower values, do not show a large enough population of low $R$ quasars that would produce a bi-modality in the distribution of $R$.

\begin{figure}
\includegraphics[width=3.5in]{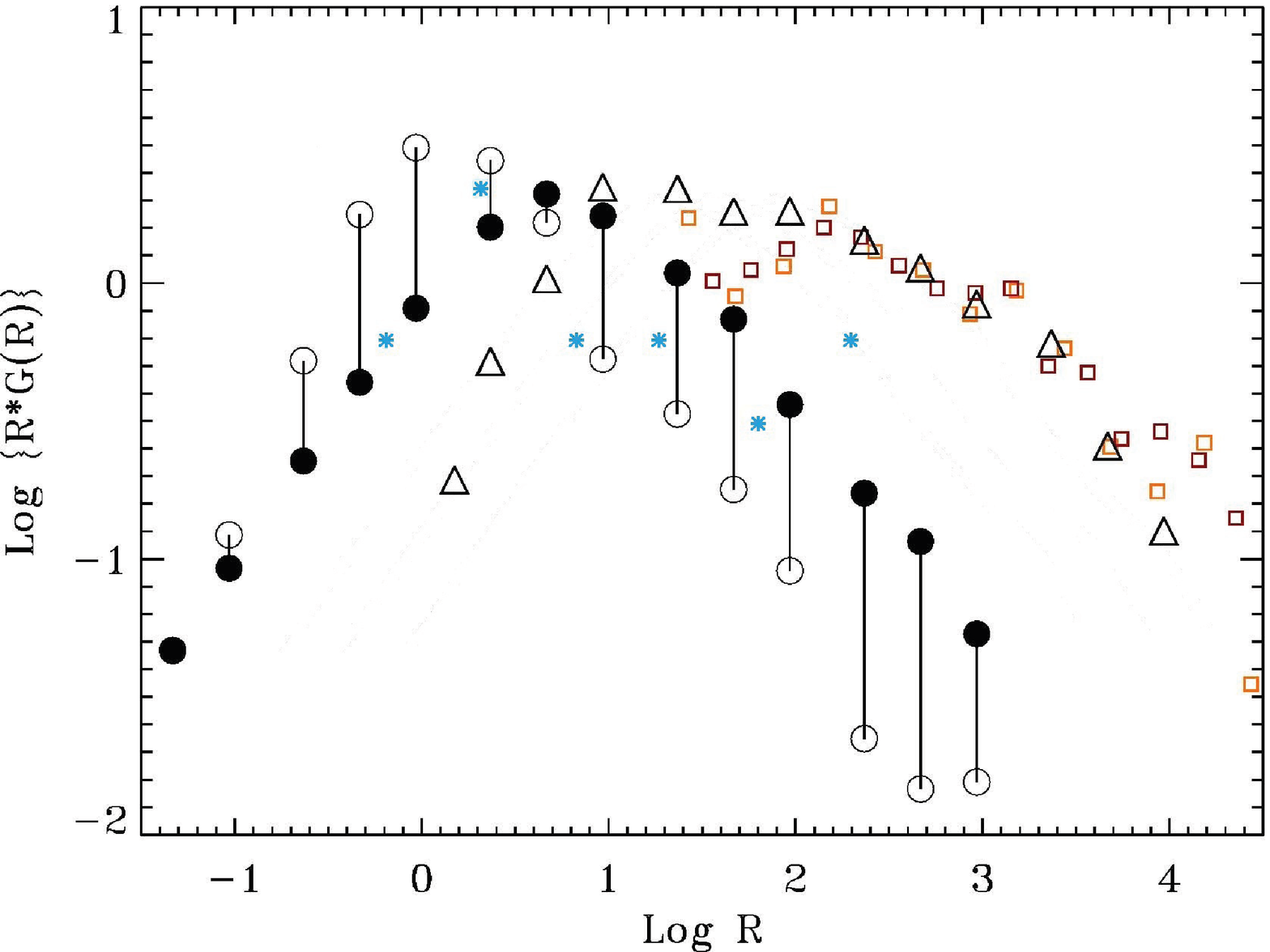}
\caption{Same as Figure \ref{psir} except without the evolved $G_R\!(R,z)$ at other redshifts, and we also overplot the raw observed $G_{R}\!(R)$ distributions obtained by \citet{Ivezic04} for $18.7< i<18$ (orange squares) and $15< i<19$ (red squares) for an SDSS $\times$ FIRST sample, and by \citet{Cira06} for a deeper radio sample which is 100\% complete to S$_{\rm 1.4 \, GHz}=60 \, \mu$Jy (blue stars).  As discussed in \S \ref{Rdist}, we would expect the distribution from \citet{Ivezic04} to agree with our raw distribution, while the distribution from \citet{Cira06} should agree with our intrinsic distribution at low values of $R$ but not necessarily high values of $R$. }
\label{psir2}
\end{figure}

In Figure \ref{psir2} we plot our results along with two determinations of the raw observed $R$ distribution available in the literature, that of \citet{Ivezic04} and that of \citet{Cira06} along with our results.  These determinations from the literature are scaled to the ratio of 5 GHz radio to 2500 \AA \, optical luminosity assuming optical and radio spectral indices of 0.5 and 0.6, respectively.  In the former case (red and orange squares on the figure), the dataset is from an SDSS $\times$ FIRST sample which is flux limited in radio and optical as in this work.  The raw $R$ distribution seen there is similar to the one seen here, although with a more pronounced dip at $R \sim$ 10.  In the later case (blue stars on the figure), the data are from a small patch of sky with deep VLA observations where the sample of SDSS quasars has 100\% completeness down to $S_{\rm 1.4 \, GHz}=60 \, \mu$Jy, and with such radio completeness we would expect the observed distribution to be consistent with our reconstructed intrinsic distribution at low values of $R$, which indeed it is, but not necessarily at high values of $R$, due either to cosmic variance or because they have significant radio flux but insufficient optical flux to be included in the SDSS sample (see the discussion in \S \ref{disc}).

\section{Tests of assumptions and error analysis}\label{testass}

{\it Luminosity dependent density evolution}:  One may be concerned that luminosity dependent density evolution (LDDE), which is not considered in the functional forms for the LF used here, may be necessary to represent the evolution of the LF.  As a test of this scenario, we divide the data into high and low halves of de-evolved luminosity $L'$ (cutting on optical luminosity), and check the similarity of the computed density evolutions for the two sets.  The density evolutions computed for both cuts are similar to each other, with the high half peaking in $\rho(z)$ at $z=1.3$ and the low redshift half peaking in $\rho(z)$ at $z=1.1$.  Given the similarity of these distributions to each other and to that computed from the dataset as a whole, we conclude that we are justified in neglecting LDDE for the purposes of this analysis.

{\it Optical measurement errors}:  There is the possibility that errors in the optical magnitudes could lead to a bias which could affect the results \citep{Eddington40}.  The bias introduced by these errors would be negligible if the source counts as a function of flux, i.e. the $\log N - \log S$ distribution, was flat (i.e. $N(>S) \propto S^0$ and $dN/dS \propto S^{-1}$). Since the number density of sources increases with decreasing flux, it is more likely that a source will be included than excluded.  However, the magnitude of this effect will depend on the faint end source counts slope, and the shallower the slope the smaller the effect.  For constant fractional flux measurement errors, an error will be introduced on the normalization of the source counts, and therefore the luminosity functions, and can be approximated by [$1/2 \, \delta^2 \, m_{below} \, (m_{below}+1)$] \citep{Teer04} where $\delta$ is the fractional error in flux and $m_{below}$ is the faint end differential source counts power law slope.  To be conservative, we will adopt errors of 0.2 magnitudes in $i$ band, to account for the intrinsic RMS scatter due to source variability, although the typical reported SDSS DR7Q measurement errors are lower, on the order of a few hundredths of a magnitude.  For the faint end magnitudes of 19.1 and this error, $\delta$ is around 0.16.  For $m_{below} \sim 2$ the bias on the luminosity function normalization will be around 7.6\%.  On the other hand, there will be an effect on the reconstructed slope of the luminosity function only if the fractional measurement errors change with luminosity.  The data show only a modest dependence on $i$ band reported error with magnitude at magnitudes below 19.1, with magnitude errors at most a factor of 1.5 higher at the highest magnitude end of that range than for the lowest magnitudes, which corresponds to a small fractional flux error.  Because data at a wide range of luminosities corresponds to a given flux, we consider this effect to be negligible, even if scatter due to variability has some flux dependence.  

{\it Radio incompleteness}: Lastly, the selection function for the FIRST objects used here might not be a sharp Heaviside function at a peak pixel flux density of 1 mJy, but rather smeared out.  According to Figure 1 of J07, the selection function of FIRST for SDSS optically identified quasars in SDSS data release 3 is such that at an integrated flux density of 1 mJy only about 55\% of sources are seen, and this number rises to 75\% at 1.5 mJy and about 85\% at 2 mJy.\footnote{The fuzzyness of the truncation boundary has a similar effect as the data measurement errors in the sense that it is unimportant for $m_{below}=-1$ and more important for larger deviations from this value.}  We have considered the sample to be limited by the peak pixel flux (i.e. surface brightness limited) in the radio rather than being limited by the integrated flux, in accordance with the criteria set forward in \citet{FIRST1}.  

The way to test the effects on our analysis of possible radio incompleteness is to repeat the analysis limiting the sample to a higher radio flux, where the sample would presumably be more complete, and determine the extent to which the calculated parameters change in a systematic way.  We have done so with lower radio flux limit of 2 mJy (4251 objects in the canonical sample), as opposed to the original 1 mJy (5445 objects in the canonical sample).  The effect, propagated through the analysis, is quite minor.   The 1$\sigma$ uncertainties on $k_{\rm opt}$  and $k_{\rm rad}$ are extended slightly.  There is no discernable effect on the correlation parameter $\alpha$.  There is a negligible effect on the density evolutions, a small effect on the local luminosity functions which are reported in \S \ref{local}, and a small effect on the shape of the $G_{R}\!(R)$ distribution which is sub-dominant to the effect of considering the radio-optical luminosity correlation to be intrinsic or induced.  To the extent that faint flux radio incompleteness is present in the sample considered here, it does not seem to have a large systematic effect on the determination of the parameters in, and conclusions of, this analysis.

\section{Discussion}\label{disc}

We have used a general and robust method to determine the radio and optical luminosity evolutions simultaneously for the quasars in the SDSS DR7 QSO $\times$ FIRST dataset, which combines 1.4 GHz radio and $i$-band optical data for thousands of quasars ranging in redshifts from 0.64 to 4.82 and over five orders of magnitude in radio loudness.

As can be seen, the results for the SDSS DR7 QSO $\times$ FIRST dataset employed here are similar in bulk to the results for the White et al. dataset that we presented in QP1.  In this work we show that, importantly, the major conclusions are not sensitive to whether the correlation between the radio and optical luminosities is considered to be induced by similar redshift evolution or intrinsic.  In the previous work, we assumed the correlation to be intrinsic.  Additionally, the luminosity evolution found for the optical only dataset matches that found for the combined radio-optical dataset, providing a test that the technique has properly handled the truncations and correlations inherent in the data.  

We have seen that the results of this analysis are nearly identical for the two different radio-optical matching radius criteria considered, 1) that adopted by J07 with a $5''$ radius for single matches and a $30''$ radius for multiple matches, and 2) a universal $5''$ matching radius which we feel is more appropriate.  The reason for this preference is that $30''$ radius corresponds to a physical scale $> 100$\,kpc \emph{projected}, at any redshift $z>0.2$.  Extended lobes in radio galaxies are known to reach similar or even larger sizes. These are however objects viewed at large inclinations, unlike quasars observed, by definition, at much smaller viewing angles \citep{Barthel89}. With such small viewing angles, projection effects are expected to limit the projected sizes of the extended structures in radio-loud quasars.  In addition, evolutionary effects may play a role as well, with distant quasars being typically younger, and therefore more compact in radio, than nearby evolved radio galaxies. This is supported by a detailed investigation of radio morphologies of SDSS (DR3) $\times$ FIRST quasars by \citet{dev06}, who noted that overwhelming majority of the selected sources are characterized by compact radio morphologies. For all these reasons, we believe that applying a universal $5''$ matching radius criterion in assigning FIRST counterparts to the SDSS bona fide \emph{quasar} sources is more appropriate, while $30''$ criterion for multiple matches has a high probability of including radio flux from objects unrelated to the optical quasar that they are supposedly linked to.

\subsection{Differential evolutions toward radio loudness with increasing redshift}\label{rldisc}

Quasars are seen to exhibit positive luminosity evolution in both wavebands, with stronger positive evolution in radio than optical.  The differential evolution can be seen in the evolution of $G_{R'}\!(R')$ as shown in Figure \ref{psir}.  This is in agreement with the results presented for a combined FIRSTxPOSS-I dataset in QP1, but in likely disagreement with J07 which claimed that the fraction of RL quasars decreases with increasing redshift.  However, \citet{Miller90} have noted that the fraction of RL quasars may increase with redshift and \citet{Donoso09} reached the same conclusion by computing radio and optical LFs at different redshifts.  \citet{Cira06} also found that the RL fraction of quasars may modestly increase at high redshift, and the conclusions of \citet{Balokovic12} favor $R$ increasing with redshift.  We note that a trend toward increasing radio loudness with redshift is visible to the eye even in the raw observed data as in Figure \ref{LvsL}, although this data suffers from biases so that the true evolution of $R$ can only be recovered with an analysis method such as employed here.

We attribute our apparent disagreement with the J07 result to three causes: 1) They present one particular moment of the $R$ distribution --- the``radio loud fraction'' --- in bins, while we are presenting the distribution itself, so direct comparison is rather difficult, 2) they are calculating this fraction by including optical sources that have no radio detection, which is a potentially correctable bias, and 3) they are not including the effect of neglecting sources that are radio bright enough to appear in FIRST but don't appear in the SDSS quasar sample that they use.   

There are potentially many AGN sources in the FIRST catalog that are not present in the SDSS quasar sample, due to their low optical fluxes.  The sample used in J07 to calculate the radio loud fraction in bins would be missing this population of sources with detectable radio emission but lower optical fluxes than would allow an SDSS detection or classification as a quasar source, while we have used an analysis which takes both flux limits properly into account.  While for the joint dataset the present work has used only quasars detected in both radio and optical and including both optical-color-selected and radio-match-selected sources, J07 used as their sample quasars detected in optical regardless of a radio detection and their quasars were selected based solely on optical color.  

Stepping back, there are four basic options when constructing a combined optical-radio dataset of quasars for an analysis of the radio loudness distribution and its evolution, given that many more confirmed quasars have an optical detection than a radio one:

I. quasars detected in both radio and optical

II. quasars detected in optical regardless of a radio detection

A. optical quasar candidates selected for followup based on either optical colors or presence of a radio match

B. optical quasar candidates selected for followup based only on optical colors

Under the above rubric, we have used IA for the joint radio-optical dataset, while J07 used IIB.

We believe that while using option II is appropriate to investigate the optical luminosity function and its evolution, such a dataset is not appropriate for investigating the radio loudness distribution and its evolution.  This is because with such a dataset one will always be introducing a bias by not including objects bright in radio but dim in optical, and at the same time, not being able to say if such a population exist at all by means of some censored data method.  The situation could be different if one were able to select two quasar samples based separately on the radio and optical data, down to given radio and optical flux limits of the surveys, and then to match the two samples forming a master-list with both radio and optical fluxes provided for each object included.  Only then non-detections in either radio or optical bands could be considered, and the censored data method could be applied \citep[see in this context, e.g.,][]{Feigelson83,Feigelson85,Isobe86}.  However since an optical identification is needed to claim a quasar nature of a source the first place, this is not possible, so in using an only optically detected set one will be always biased against a radio-loud population.  On the other hand, if, as in the present work, one uses option I, and then analyzes the data in such a way as to take the flux limits into account, an unbiased reconstruction of the intrinsic properties of the population can be achieved. 

The question of whether option A or B is most appropriate is less straightforward.  However, quasar optical color may be correlated with radio loudness \citep{Kimball11}.  If this is the case, then by selecting quasars based on the optical colors solely, one may introduce some bias when dealing with the radio loudness distribution and evolution. Option IA means that the color criterion is not important: by including only those quasars which are detected in radio, and by allowing for radio-selected optical quasar candidates, we basically end up with a radio-selected (and not color-selected) and radio and optical flux limited sample of quasars.

\subsection{Lack of bi-modality in radio loudness}\label{lbmdisc}

Our results favor a single population of quasars with no bi-modality in the radio loudness parameter for the range of $R$ considered here, in agreement with the conclusions of QP1.  Some physical implications of a single population in this range of $R$ are discussed in that work, including that there is thus no evidence for the existence of two physically different population of quasars with respect to the radio/jet production efficiency.  Although it is a somewhat different issue, there have also been studies of radio populations investigating the distribution of quasars, Seyferts, and other types of AGN in the plane of radio loudess versus accretion rate (note that by definition all the quasars accrete at high and very high rates, unlike radio galaxies or Seyferts). These results indicate that there may be two `tracks' in the space, characterized by a similar non-monotonic dependence of the radio loundess parameter on the Eddington ratio, but differing in normalization, and corresponding to what might be considered as a RL population hosted exclusively by elliptical galaxies, and a RQ population of AGN hosted by both types of galaxies \citep{Sikora07}. Such studies were however done --- and in fact had to be done --- on incomplete and inhomogeneous samples of AGN, and were claimed to depend on whether core or total radio luminosities are used \citep[e.g][]{BF11}. The host-related bimodal distribution of radio loudness in the entire AGN population is however distinct from the main problem addressed in this work, dealing strictly with a well-defined population of quasar-type AGN (for which, notably, we use the FIRST --- i.e. the total --- radio fluxes). And in particular, the conclusion that there is no radio bimodiality for the quasar population does not contradict the findings presented in \citet{Sikora07}, as these authors emphasized that while a complete sample of elliptical-hosted AGN (including quasars) is expected to show continuous distribution of the radio loudness down to the ``radio quiet'' regime, the point is that the spiral-hosted AGN do never reach high values of the radio-loudness parameters (i.e., values comparable to those of radio galaxies), even though at very low accretion rates they are often characterized by $\log R >1$ (see section 3 of that work and the discussion therein).

In \S \ref{Rdist} we explored additional arguments against the presence of a bi-modality in $R$ at lower values of $R$, below those considered in this analysis. The caution to the above conclusion is that the sample analyzed here is dominated by RL objects, as we do not, for the reasons discussed above, consider SDSS quasars with no detected FIRST counterparts.  Figure \ref{LvsL} indicates that with this SDSS $\times$ FIRST dataset we are sensitive to regions where additional RQ objects would lie were they to exist (e.g. below the $R=10$ line and above the $L_{rad,min}$ lines in that figure) but there does not appear to be an overabundance of those objects.   However, we cannot formally exclude the possibility that there may be a population of very radio quiet quasars which would form a separate branch in Figure \ref{psir} were the abscissa extended to the left, resulting in a bi-modal distribution of radio loudness for the entire quasar population. 

Such a numerous population of very radio-quiet quasars, even if absent in the FIRST database, would not be missed in deeper radio surveys probing sub-mJy flux levels.  In fact, \citet{Kimball11} have used EVLA observations to detect 179 SDSS DR7Q quasars in the range $0.2<z<3$ in their field down to a 6 GHz flux limit of 20 $\mu$Jy.  They claim to detect almost every quasar (97\%) in the field in radio, and construct the 6 GHz radio luminosity function with the $V/V_{\rm max}$ method.  Their radio luminosity function is well fit with two populations, one with radio emssion dominated by AGN central engine emsision, and one with radio dominated by star formation.  The later becomes prominent at luminosities below $L_{\rm 6 GHz} \sim 10^{30}$ erg s$^{-1}$ Hz$^{-1}$.  This corresponds to radio luminosities just below those probed in this analysis, as can be seen in Figure \ref{psirad}, and would imply a two population model with the more RQ population dominating at $R \sim 1$ and below, although without a 'dip' at $R \sim 10$.  

\citet{Mahony12} claim no bimodality in the $R_{20}$ radio loudness parameter with X-ray selected sample with $z<1$ down to $\sim 0.2$ mJy.  The \citet{Kimball11} result disfavors a large population with values of $R$ below $R<0.1$ that would cause a bi-modality in $R$, based on a volume-limited sample probed to low radio fluxes.  Both of these works claim evidence for a low radio flux population of quasars where the radio emission is significantly enhanced or even dominated by star formation activity in the host galaxies.  This is largely a different issue than the ``traditional'' bimodality question in quasars, i.e. as to whether there is a two-component shape to the radio loudness distribution centered around $R=10$.

The measured $dN/dS$ distribution of extragalactic radio sources at low fluxes, together with the limiting level of the extragalactic radio background \citep{Singal10,Fixsen11} and the quasar optical LF as constructed here, could in principle provide further constrains on the distribution of the radio-loudness parameter at very low values of $R$. This will be addressed in a forthcoming analysis.

\acknowledgments

Funding for the SDSS and SDSS-II has been provided by the Alfred P. Sloan Foundation, the Participating Institutions, the National Science Foundation, the U.S. Department of Energy, the National Aeronautics and Space Administration, the Japanese Monbukagakusho, the Max Planck Society, and the Higher Education Funding Council for England. The SDSS Web Site is http://www.sdss.org/.

\clearpage

\appendix

\section*{Notes on the effects of correlated radio and optical luminosities for this analysis}

As discussed in \S \ref{ifcorr}, in the case of intrinsically correlated luminosities, the luminosity functions in different wavebands are not a-priori separable and one must do a coordinate rotation of the form in equation \ref{rcrdef} to a correlation reduced luminosity to achieve a separable luminosity function.  In that case, one must determine the two now independent luminosity-redshift correlation functions $g_{\rm opt}\!(z)$ and $g_{\rm crr}\!(z)$ which describe the luminosity evolutions.  The full procedure is detailed in \S \ref{evsec}, and this result can be compared to the evolutions determined from assuming no intrinsic (i.e. only redshift induced) correlation between the radio and optical luminosities, to determine whether the effect is significant.  

The local LFs of uncorrelated luminosities $L'_{\rm opt}$ and $L'_{\rm crr}$ can then be used to recover the local radio LF by a straight forward integration over $L'_{\rm crr}$ and the true local optical LF as 

\begin{eqnarray}
\psi_{\rm rad}\!(L_{\rm rad}') = 
\nonumber \\ \int_0^{\infty} { \psi_{\rm opt}\!(L_{\rm opt}') \, \psi_{\rm crr}\left({{ L_{\rm rad}' } \over ({{L_{\rm opt}'/L_{\rm fid}})^{\alpha} } } \right) \, {{dL_{\rm opt}'} \over ({{L_{\rm opt}'/L_{\rm fid}})^{\alpha}}  } \,}
\label{localrad}
\end{eqnarray}
As stated above this procedure can be used for the determination of the radio LF at any redshift, from which one can deduce that the radio luminosities also undergo luminosity evolution with 
\begin{equation}
g_{\rm rad}\!(z) = g_{\rm crr}\!(z) \, \times \, [g_{\rm opt}\!(z)]^{\alpha}
\label{gradform}
\end{equation}
(cf equation \ref{rcrdef})

Similarly we can determine the local distribution of the radio to optical luminosity ratio, 
$R' = L'_{\rm rad}/L'_{\rm opt} = L'_{\rm crr} \, \times \, {L'_{\rm opt}}^{\alpha-1} \, \times {L_{\rm fid}}^{-\alpha}$, as

\begin{equation}
G_{R'} = \int_0^{\infty} { \psi_{\rm opt}\!(L_{\rm opt}') \, \psi_{\rm crr}\left({{ R' \, L_{\rm fid}} \over ({{L_{\rm opt}'/L_{\rm fid}})^{\alpha-1} } } \right) \, {{dL_{\rm opt}'} \over {{L_{\rm opt}'}^{\alpha - 1} \, L_{\rm fid}}  } \,}
\label{localr}
\end{equation}
and its evolution 
\begin{equation}
g_{R}\!(z) = g_{\rm crr}\!(z) \, \times \, [g_{\rm opt}\!(z)]^{\alpha-1} = {g_{\rm rad} \over g_{\rm opt}}
\label{Rexp}
\end{equation}

These can then be compared with $\psi_{\rm rad}\!(L_{\rm rad}')$ and $G_{R'}$ as determined from assuming no intrinsic correlation between the radio and optical luminosities, which is done in Figure \ref{psirad} and Figure \ref{psir}.  We see that the main conclusions of this work are not dependent on whether the observed correlation between $L_{\rm rad}$ and  $L_{\rm opt}$ is intrinsic or induced.  

\end{document}